\documentclass[%
 reprint,
superscriptaddress,
msmath,amssymb,
amsfonts,amsmath,floatfix
aps,
prb,
]{revtex4-1}
\pdfoutput=1
\usepackage{graphicx}
\usepackage{amsmath}
\usepackage{amssymb}
\usepackage{dsfont}
\usepackage{tensor}
\usepackage{color}
\usepackage{soul}
\usepackage{ulem}
\usepackage{cancel}
\usepackage{upgreek}
\usepackage{braket}
\usepackage{hyperref}
\usepackage[letterpaper,left=1.5cm,right=1.5cm,top=1.6 cm,bottom=2.5cm]{geometry}
\setstcolor{red}

\begin{document}
\pdfoutput=1
\preprint{APS/123-QED}

%\title{Non-Markovian dynamics induced by phonons in color centers in diamond}

%\title{A measure for non-Markovianity based on  and its applications to color centers in diamond}
\title{Coherence as a measure for non-Markovianity and its applications to color centers in diamond}

\author{Ariel Norambuena}
\email{ariel.norambuena@umayor.cl}
\affiliation{Centro de \'Optica e Informaci\'on Cu\'antica, Facultad de Ciencias, Universidad Mayor, Chile}
\author{Jer\'onimo R. Maze}
\affiliation{Instituto de F\'{\i}sica, Pontificia Universidad Cat\'{o}lica de Chile, Casilla 306, Santiago, Chile}
\author{Peter Rabl}
\affiliation{Vienna Center for Quantum Science and Technology, Atominstitut, TU Wien, 1040 Vienna, Austria}
\author{ Ra\'ul Coto}
\email{raul.coto@umayor.cl}
\affiliation{Centro de \'Optica e Informaci\'on Cu\'antica, Facultad de Ciencias, Universidad Mayor, Chile}

\date{\today}% It is always \today, today,

\begin{abstract}
The degree of non-Markovianity of a continuous bath can be quantified by means of the coherence. This simple measure is experimentally accessible through Ramsey spectroscopy, but it is limited to incoherent dynamical maps. We propose an extension of this measure and discuss its application to color centers in diamond, where the optical coherence between two orbital states is affected by interactions with a structured phonon bath. By taking realistic phonon spectral density functions into account, we show that this measure is well-behaved at arbitrary temperatures and that it provides additional insights about how non-Markoviantiy is affected by the presence of both bulk and quasi-localized phonon modes. Importantly, with only a little overhead the measure can be adapted to eliminate the false signs of non-Markovianity from coherent dynamical maps and is thus applicable for a large class of systems modeled by the spin-boson Hamiltonian.  
\end{abstract}

%\begin{description}
%%\item[References]	
%%Secondary publications and information retrieval purposes.
%%\item[Structure]
%%You may use the \texttt{description} environment to structure your abstract;
%%use the optional argument of the \verb+\item+ command to give the category of each item. 
%\end{description}

\maketitle

\section{Introduction}
By employing the principles of superposition and entanglement, quantum systems can outperform their classical counterparts in many applications such as computation, cryptography and high-precision measurements~\cite{Nielsen,Giovannetti,Gross}. However, to benefit from this quantum advantage, the systems must be protected from detrimental interactions with the environment using passive isolation as well as active techniques such as error correction or decoupling pulses. To implement efficient error-mitigation schemes it is crucial to have a precise understanding of the underlying system-environment interaction~\cite{Breuerbook}, in particular in realistic non-Markovian settings, where information can flow back from the bath to the system and is not immediately lost. These scenarios have been exploited, for example, for quantum metrology~\cite{Chin}, quantum channels~\cite{Bylicka} or quantum control~\cite{Reich} and led to a growing interest in the questions how different environments can be compared and how non-Markovianity (NM) can be quantified~\cite{Breuer,Luo,Rivas}. However, most of the proposed measures for NM so far are based on rather tedious mathematical constructs, such as the maximization of the trace distance between two different initial states~\cite{Breuer}, the addition of auxiliary systems~\cite{Luo} or the complete knowledge of the dynamical map~\cite{Rivas}. This makes proof-of-concept demonstrations and the broad use of such measures for modeling realistic applications very cumbersome. \par

In this work, we focus on a simple measure for NM, which can be directly derived from the coherence of the system of interest~\cite{Chanda,Passos}. This quantity is most essential for quantum technology applications and in many cases it can be accessed through Ramsey measurements. To demonstrate its suitability for characterizing realistic system-bath interactions in a physically meaningful way, we discuss in more detail the application of this measure for color centers in diamond. Color centers such as the negatively charged silicon-vacancy (SiV$^-$) and nitrogen-vacancy (NV$^-$) center have attracted wide attention because they can be initialized, controlled and read out with high fidelities~\cite{Gruber,Oort,Becker}. These unique properties make them strong candidates for various quantum sensing and quantum information processing application~\cite{Jero2008, Dutt2007}, but their optical properties are still limited by unavoidable interactions with phonons. The influence of a continuum of bulk modes and distinct quasi-localized resonances originates a rich and complex dynamics arising from different NM behavior beyond the extensively studied Ohmic environment \cite{Legget1987}. A deeper understanding of the NM dynamics induced by realistic phononic baths will be important for engineering and optimizing defect-phonon interactions in structured reservoirs such as 
cantilevers~\cite{Kepesidis2016}, two-dimensional layers~\cite{Abdi2019}, phonon waveguides~\cite{Lemonde2018} or phononic crystals~\cite{Mark2018,Nori2019}. \par

%Moreover, in deriving our results we provide a thorough analysis of the spectral density function that phenomenologically describe the electron-phonon interaction of an individual SiV$^{-}$ center in diamond as well as the corresponding dephasing rate. Furthermore, we compare our measure with well stablished measures of NM in the presence of thermal effects and with contribution of acoustic and quasi-localized phonon modes. Finally, we show our results are valid in the weak and strong coupling regime of the generalized spin-boson model.

\section{Measure of non-Markovianity}

One natural approach towards witnessing NM is to examine the back-flow of quantum information in terms of the 
Coherence~\cite{LoFranco}. For a given complete set of basis states $\{|i\rangle\}$, Coherence is commonly defined as $C(t)=\sum_{i\neq j}|\rho_{ij}(t)|$~\cite{Baumgratz}, where the $\rho_{ij}(t)=\langle i|\rho(t) |j\rangle$ are the matrix elements of the system density operator.  Although this definition is not unique, the choice of basis states usually follows naturally from the context and is very often taken as the eigenbasis of the bare system. For a two-level system the definition above reduces to $C(t)=\sqrt{\langle \sigma_x \rangle^2 + \langle \sigma_y \rangle^2}$, where the $\sigma_k$ are the usual Pauli operators.  In this case, $C(t)$ can be experimentally attained through standard Ramsey spectroscopy, i.e., by applying two $\pi/2$-pulses separated by a time $t$. A measurement of the final population in the excited state, $P_e = (1 + \langle \sigma_x\rangle)/2$, then provides a measurement of $\langle \sigma_x\rangle$, 
or $\langle \sigma_y\rangle$, if an additional rotation between the two pulses is introduced. Similar strategies can also be applied to measure $C(t)$ for higher dimensional systems, where, however, the pulse sequences are slightly more involved. \par

The Coherence does not increase under incoherent completely positive and trace preserving maps ($\lbrace\Lambda_{ICPTP}\rbrace$)~\cite{Baumgratz}, in other words, it is a monotonically decreasing function in time $C_{\rho}(t)\geq C_{\Lambda_{ICPTP}(\rho)}(t)$. Hence, considering a free evolution that obeys $\Lambda_{t}\in\Lambda_{ICPTP}$, one can build up a measure of NM based on the oscillations in $C(t)$, that are triggered by the coupling to an external reservoir~\cite{Chanda,Passos}. The measure of NM is then defined as
\begin{equation}\label{NM_measure}
\mathcal{N}_C(T) = 1+ \frac{\displaystyle{\int_0^{\infty}  \dot{C}(\tau) \,d\tau}}{\displaystyle{\int_0^{\infty}  |\dot{C}(\tau)| \,d\tau}}, \quad  0 \leq \mathcal{N}_C(T) \leq 1.
\end{equation}
Even though $\mathcal{N}_C(T)$ is only valid in the presence of incoherent dynamical maps, we show in Sec.~\ref{NM-Coherent-Maps} that this constraint can be relaxed and that the measure can be extended to coherent maps, with
practical applications on the generalized spin-boson Hamiltonian.

\section{Optical coherence in color centers} \label{Opt}

In this section, we investigate in more detail the application of the measure defined in Eq.~\eqref{NM_measure} to the case of SiV$^{-}$ and NV$^{-}$ centers in diamond. By taking only a single ground state $|g\rangle$ and a single electronically excited state $|e\rangle$ into account, the optical properties of the center in the presence of a phonon reservoir are well-described by the spin-boson Hamiltonian. In the frame rotating with the frequency $\omega_L$ of the driving laser, the Hamiltonian reads ($\hbar = 1$)~\cite{Legget,Weiss}
\begin{equation}
H  = -{\Delta \over 2} \sigma_z + {\Omega \over 2} \sigma_x + \sum_{k} \omega_k a_{k}^{\dagger}a_k + \sigma_z\sum_{k}g_k  \left(a_k + a^{\dagger}_k \right), \label{Hamiltonian_Spin_Boson_Model}
\end{equation}
where $\Delta=\omega_L-\omega_{eg}$ is the detuning from the bare transition frequency $\omega_{eg}$ and $\Omega$ is the optical Rabi frequency. In Eq.~\eqref{Hamiltonian_Spin_Boson_Model}, the $\omega_k$ are the phonon frequencies and  $a_k^{\dagger}$ ($a_k$) the corresponding creation (annihilation) operators for the phonon modes. The  $g_k = \lambda_{e,k}-\lambda_{g,k}$~\cite{Ariel2016} denote the effective electron-phonon coupling constants, which arise from the different deformation potentials, $\lambda_{e,k}$ and $\lambda_{g,k}$, in the ground and the excited state. \par

During the free evolution time $(\Omega=0)$ and assuming that the phonon reservoir is initially in a thermal state $\rho_{ph}$, the dynamics of the reduced color center state is described by the following time local master 
equation~\cite{Luczka,Haikka}
\begin{equation}\label{ME}
{d \rho_s \over dt} = -\frac{\gamma(t)}{2}(\rho_s(t)-\sigma_z\rho_s(t)\sigma_z),
\end{equation}    
where we have omitted a Hamiltonian contribution $\sim [\sigma_z, \rho_s(t)]$, which does not affect the coherence. Therefore, the system-environment interaction is fully determined by the time-dependent dephasing rate~\cite{Luczka} ($\hbar = 1$)
\begin{equation} \label{DephasingRate}
\gamma(t) = \int_{0}^{\infty} {J(\omega) \over \omega} \coth\left({\omega \over 2 k_B T}\right)\sin(\omega t) \, d\omega,
\end{equation}
where $J(\omega) = \sum_{k} |g_k|^2\delta(\omega - \omega_k)$ is the spectral density function (SDF), $k_B$ is the Boltzmann constant and $T$ is the reservoir temperature. In general, the SDF $J(\omega)$ satisfies two important properties: i) $J(0) = J(\omega>\omega_{max})=0$ and ii) $J(\omega) > 0 \; \forall \; \omega \in (0, \omega_{max})$, where $\omega_{max}$ is the largest phonon frequency of the reservoir. The formal solution of the off-diagonal elements of $\rho_s(t)$ is given by $\rho_{eg}(t) = \rho_{eg}(0)e^{-\Gamma(t)}$, where $\Gamma(t) = 2\int_{0}^{t}\gamma(\tau) \, d\tau$ is a bounded function that satisfies $0 \leq \Gamma(t) \leq 4 \int_{0}^{\infty}J(\omega)/ \omega^2 \mbox{coth}(\omega/2k_B T)\, d\omega$. 

\begin{figure}
\centering
\includegraphics[width=0.8 \linewidth]{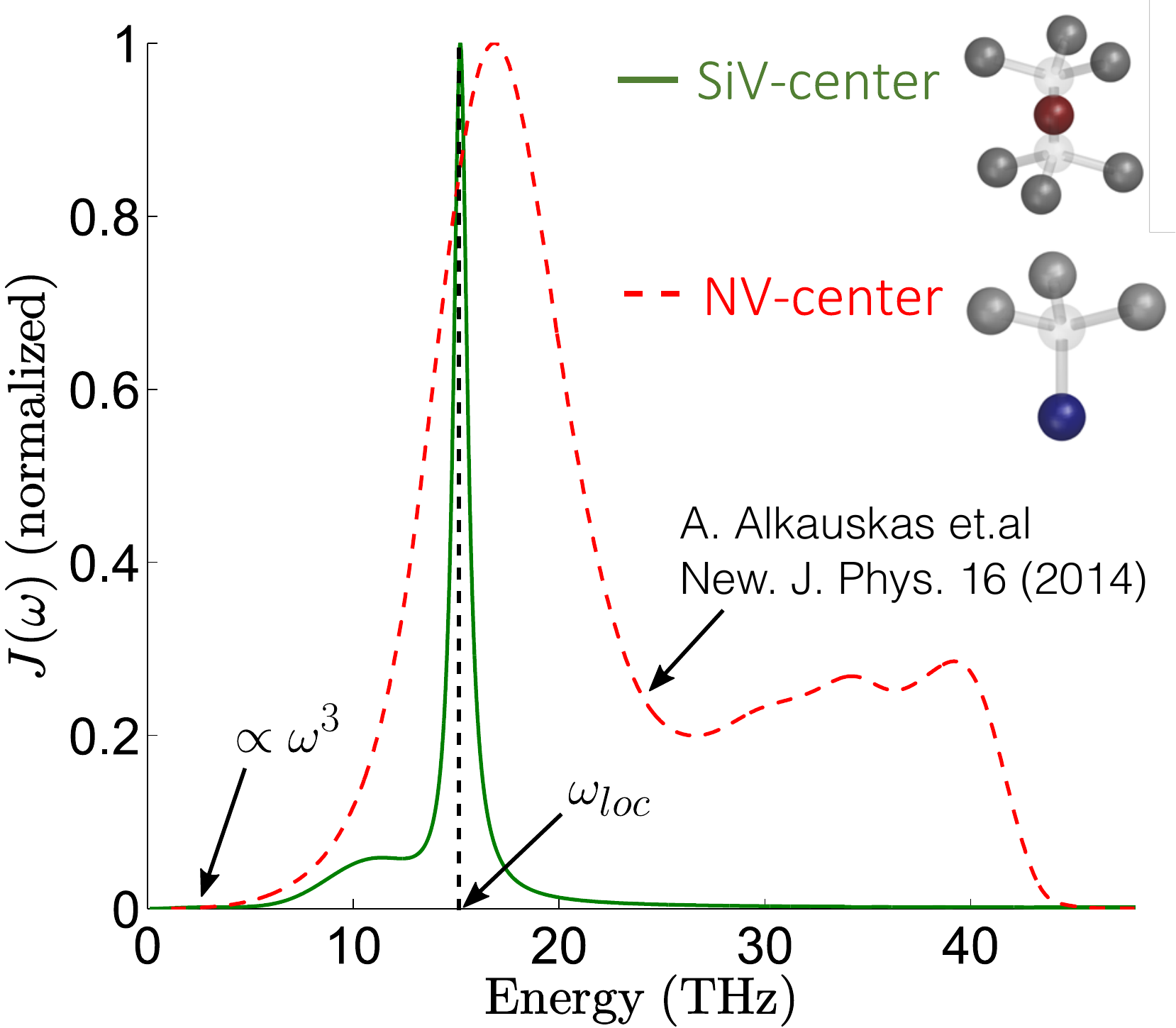}
\caption{(Color online) Spectral density function of the SiV$^{-}$ and NV$^{-}$ centers. Green (solid) and red (dashed) curves correspond to the phenomenological model $J(\omega) = J_{bulk}(\omega)+J_{loc1}(\omega)+J_{loc2}(\omega)$ given in Eqs.~\eqref{Jbulk}-\eqref{Jloc2} and the first-principles spectral density function calculated in Ref.~\cite{Alkauskas}, respectively. At low frequencies, both curves have a dominant contribution of acoustic phonons leading to $J(\omega) \propto \omega^3$. The main peak of the green curve (SiV$^{-}$) is the contribution of a quasi-localized phonon mode with frequency $\omega_{loc} \approx 15.19$ THz.}
    \label{fig:Figure1}
\end{figure} 

\section{Phonon spectral density and dephasing rate}

From Eq.~\eqref{DephasingRate} we see that the dephasing rate $\gamma(t)$, and thus the degree of NM, depends only on the SDF, $J(\omega)$, and the temperature $T$. In the case of color centers or other solid-state emitters, information about the SDF can be obtained from the photoluminescence (PL) spectrum~\cite{Dietrich2014, Ariel2016}, where the coupling to the phonons both reduces the bare resonance (zero-phonon line) and leads to additional phonon-sidebands. 
The experimental PL spectrum of the SiV$^{-}$ center exibits an isotopic shift feature in the prominent and narrow phonon sideband~\cite{Dietrich2014}. This can be explained by the strong electron-phonon interaction with a quasi-localized phonon mode primarily composed of a large oscillation of the silicon atom~\cite{Dietrich2014, Ariel2016, Londero2018}. In addition, lattice vibrations in the bulk lead to a smooth SDF, which typically scales as $\sim\omega^3$ for low frequencies in a three-dimensional lattice.  The phononic SDF that reproduces the isotopic shift feature of the PL spectrum and the effect of acoustic phonons is phenomenologically given by $J(\omega) = J_{bulk}(\omega) + J_{loc1}(\omega ) + J_{loc2}(\omega)$, with~\cite{Ariel2016}
\begin{eqnarray}
J_{bulk}(\omega) &=& 2\alpha \omega_c^{1-d} \omega^d e^{-\omega/\omega_c}, \label{Jbulk} \\
J_{loc1}(\omega) &=& {J_0 \omega^d \over \left({\omega \over \omega_{loc}}+1 \right)^{2}}\frac{\Gamma/2}{(\omega-\omega_{loc})^2 + (\Gamma/2)^2},\label{Jloc1} \\
J_{loc2}(\omega) &=& J_1 \omega^d e^{-(\omega-\omega_0)^2 /(2 \sigma^2)}, \label{Jloc2}
\end{eqnarray}    
where $d$ is the dimension of the diamond lattice ($d =3$ in our case). Acoustic phonons are associated with 
low-energy vibrational excitations where the atoms of the color center are oscillating in phase and therefore experimenting a weak electron-phonon interaction. Thus, these phonons are reasonably well described by an intensity $\alpha$ and a cutoff frequency $\omega_c \simeq 1$ THz. However, quasi-localized phonon modes induce out-of-phase oscillations of the defect's atoms with large amplitude leading to a strong electron-phonon interaction. This type of interactions is usually modeled by Lorentzian-like functions centered around the specific localized phonon frequency $\omega_{loc}$, with a characteristic width $\Gamma$ and an intensity $J_0$. For the region in between 
(from 1 THz to $\sim$14 THz) other vibrational modes participate~\cite{Gergo2018}, which is captured by the Gaussian contribution $J_{loc2}(\omega)$. For the SiV$^{-}$ center, the following parameters of the SDF are chosen to accurately match the PL spectrum obtained from molecular dynamic simulations: $\alpha = 0.0275$, $J_1 = 0.0025$ THz$^{-2}$, $\sigma = 2.4042$ THz, $\omega_0 = 9.35$ THz,  $J_0 = 0.0235$ THz$^{-1}$, $\Gamma = 0.8414$ THz, and $\omega_{loc}=15.19$ THz. For the NV$^-$ center we do not make this decomposition and use the exact SDF obtained from a detailed first-principles calculation reported in Ref.~\cite{Alkauskas}. In Figure~\ref{fig:Figure1} we plotted the normalized spectral density functions of both NV$^{-}$ and SiV$^{-}$ centers. \par

By following the same partition as for the SDF, we write for the SiV$^-$ center the total dephasing rate defined in Eq.~\eqref{DephasingRate} as $\gamma(t)=  \gamma_{bulk}(t)+\gamma_{loc1}(t)+\gamma_{loc2}(t)$. In order to better illustrate the behavior of the individual contributions we will first consider the limit of very low temperatures where $\coth(\omega/2k_BT) \approx 1$  and  
\begin{eqnarray}
\gamma_{bulk}^{\downarrow}(t,d) &=& 2\alpha\omega_c(d-1)!{\sin(d\tan^{-1}(\omega_c t)) \over \left[1+(\omega_c t)^2\right]^{d/2}}, \label{gamma_bulk_lowT}  \\
\gamma_{loc1}^{\downarrow}(t) &\approx & {1 \over 4}J_0 \omega_{loc}^2 \pi\sin(\omega_{loc}t)e^{-\Gamma t/2},\label{gamma_loc1_lowT} \\
\gamma_{loc2}^{\downarrow}(t) &=& J_1 \int_{0}^{\infty} \omega^2 e^{-(\omega-\omega_0)^2/(2\sigma^2)}\sin(\omega t)\, d\omega. \label{gamma_loc2_lowT}
\end{eqnarray}
For a detailed derivation of the dephasing rate $\gamma_{loc1}^{\downarrow}(t)$ see Appendix~\ref{Appendix_gamma}. We left $\gamma_{loc2}^{\downarrow}(t)$ in terms of the integral since it involves a rather complicated expression. 
Note that Eq.~\eqref{gamma_bulk_lowT} is only valid for $d>-1$, which is satisfied in our case ($d =3$). In the opposite, high temperature regime, $\omega/2k_BT\ll 1$ and  $\coth(\omega/2k_BT)\approx 2k_BT/\omega$. In this case we obtain $\gamma_{bulk}^{\uparrow}(t) =  \left(2 k_B T / \omega_{c} \right)  \gamma_{bulk}^{\downarrow}(t,d-1)$, $\gamma_{loc1}^{\uparrow}(t) \approx   \left(2 k_B T / \omega_{loc} \right) \gamma_{loc1}^{\downarrow}(t)$ and $\gamma_{loc2}^{\uparrow}(t) = 2k_B T J_1 \int_{0}^{\infty} \omega \mbox{exp}(-(\omega-\omega_0)^2/(2\sigma^2))\sin(\omega t)\, d\omega$. \par 

\begin{figure}[ht!]
\centering
\includegraphics[width = 1 \linewidth]{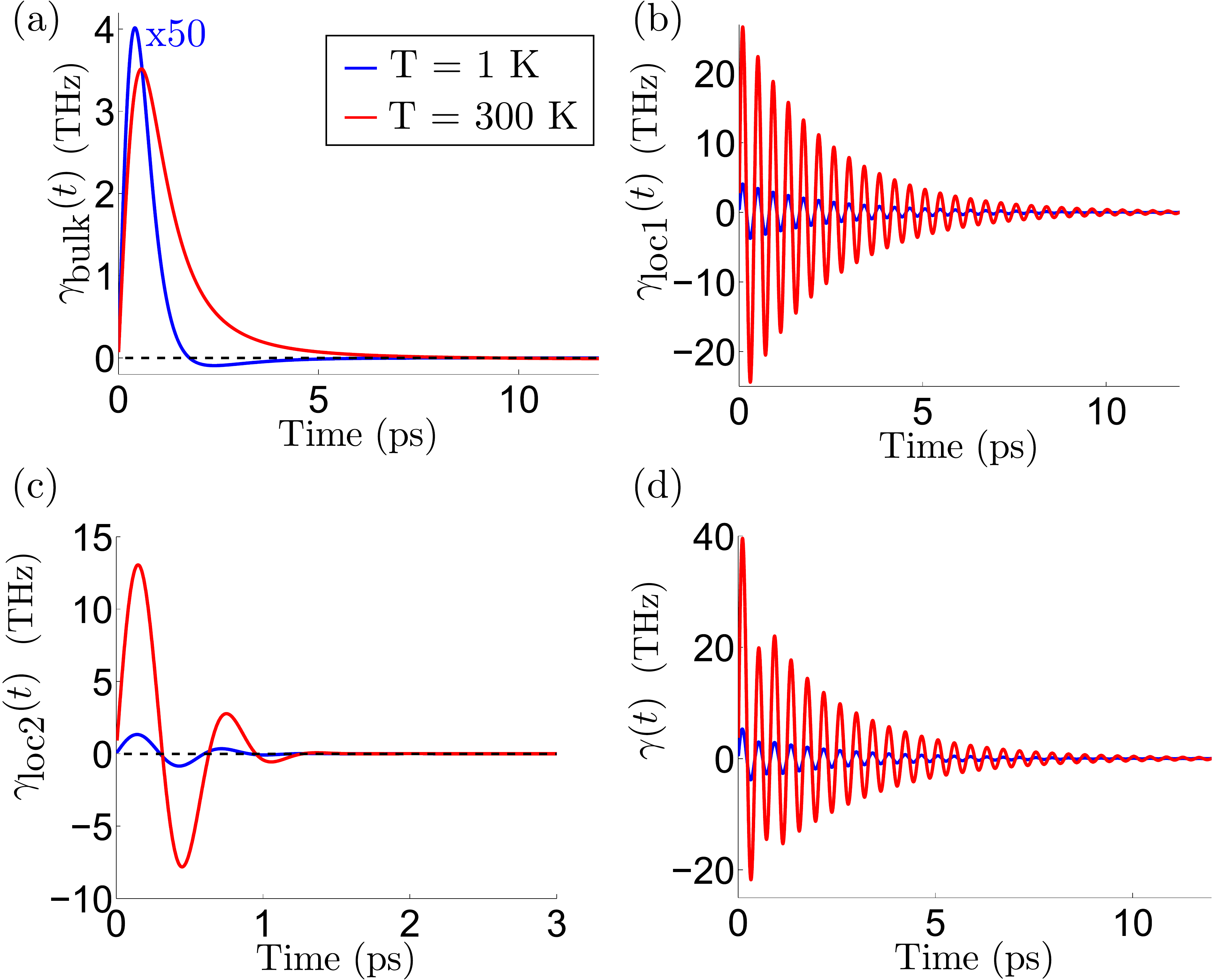}
\caption{The dephasing rate is plotted separately by considering each contribution of the phononic spectral density function: (a) $J_{bulk}(\omega)$ (acoustic phonons), (b) $J_{loc1}(\omega)$ (strong interaction with a quasi-localized phonon), (c) $J_{loc2}(\omega)$ (gaussian spectral function), and (d) the whole contribution given by $J(\omega) = J_{bulk}(\omega) + J_{loc1}(\omega) + J_{loc2}(\omega)$. Blue and red curves shows temperature effects on the dephasing rate for $T = 1$ K and $T = 300$ K, respectively. In (b) and (d) the period of the oscillations are approximately given  by $2\pi / \omega_{loc} \approx 0.41$ ps, where $\omega_{loc} = 15.19$ THz is the frequency associated with the strong electron-phonon coupling illustrated in Figure~\ref{fig:Figure1}.}
\label{fig:Figure2}
\end{figure}  

Figure~\ref{fig:Figure2} shows the expected dephasing rates for the SiV$^-$ center in diamond for low and high temperatures. Moreover, based on numerical simulations we set the low (high) temperature regime to be below (above) $T \approx 20$ K ($286$ K). These boundaries have been chosen such that the approximate low- and high-temperature limits for the dephasing rate given above match its exact numerical shape. From the statistical properties of the phonon reservoir, in particular, the mean number of phonons $n(w) = [\mbox{exp}\left(\hbar \omega /k_B T \right)-1]^{-1}$, we observe that $n(\omega_{loc})\approx 2$ for $T \approx 286$ K and $\omega_{loc} = 15.19$ THz. In other words, the high-temperature regime is defined from the thermal activation of the strong electron-phonon coupling with the quasi-localized phonon mode. \par 

If we individually look at the components of $\gamma(t)$, it is straightforward to notice that each one will get involved in the dynamics at a different temperature. For $\gamma_{bulk}(t)$ in Figure~\ref{fig:Figure2}-(a) we observe a region where it takes negative values, and above a critical temperature ($T_c \approx 1.85$ K) it is always positive. However, $\gamma_{loc1}(t)$ and $\gamma_{loc2}(t)$ show negative values even at room temperature, with an amplitude that is non-negligible as compared to the low-temperature case, see Figures~\ref{fig:Figure2}-(b) and \ref{fig:Figure2}-(c). This negative behavior has previously been connected to memory effects \cite{Luo}, as we will detail in what follows. Finally, one can notice that $\gamma_{loc1}(t)$ has the leading contribution to the dephasing rate, see 
Figure~\ref{fig:Figure2}-(d). For comparison, in Appendix~\ref{Appendix_NV} we also evaluate the dephasing rate for an NV$^{-}$ center, which also exhibits negative values at room temperature. \par
  
\section{Non-markovianity in color centers}
   
Let us now discuss the degree of NM of the described phonon environment, as quantified through $N_C(T)$. To begin, we first introduce another related measure for NM, which we will use for comparison. Following the procedure in Ref.~\cite{Hall}, where the figure of merit is the canonical decay rate $\gamma^c(t)$ when the Lindbladian is written in an orthonormal basis for a $d$-level system $\mathcal{L}[\rho(t)]=-i[H(t),\rho(t)] + \sum_{j=1}^{d^2-1}\gamma_j^c(t)( L_j(t)\rho(t)L_j^\dagger(t) - \lbrace L_j^\dagger(t)L_j(t),\rho(t)\rbrace /2)$, where $\mbox{Tr}[ L_i^\dagger(t)L_j(t)]=\delta_{ij}$. Based on this rate, the function $f(t)\equiv\max\lbrace -\gamma^c(t), 0\rbrace=(\vert\gamma^c (t) \vert -\gamma^c (t))/2$~\cite{Hall} is introduced. For our particular case where the effect of the environment represented in Eq.~\eqref{ME} only induces a pure dephasing dynamics and the operators involved form an orthogonal basis, we have $\gamma^c(t) = 2\gamma(t)$~\cite{Hall,Rivas}. Therefore, a NM measure ($\mathcal{N}_{\gamma}$) as a function of the reservoir temperature is defined when integrating $f(\tau)$ over a bound time interval $\mathcal{N}_{\gamma}(T)  = \int_{t}^{t'} \left(\vert\gamma (\tau) \vert -\gamma (\tau)\right)\, d\tau$~\cite{Hall,Rivas}. At first glance, we can witness NM from the negative values of $\gamma(t)$~\cite{Luo,Vacchini,Haikka,Rivas,Breuer}, which means that the previous discussion about the negative behavior of $\gamma(t)$ in Figure~\ref{fig:Figure2} stands as a proof of NM for the orbital states of the SiV$^-$ center,--- similar conclusions are obtained for the NV$^-$ center. This result is the first evidence that the phononic contribution induces NM behavior in color centers in diamond, commonly modeled as purely Markovian~\cite{Ralf2014, Kepesidis2016, Ariel2018}. \par

\begin{figure}[ht!]
\centering
\includegraphics[width = 1.01 \linewidth]{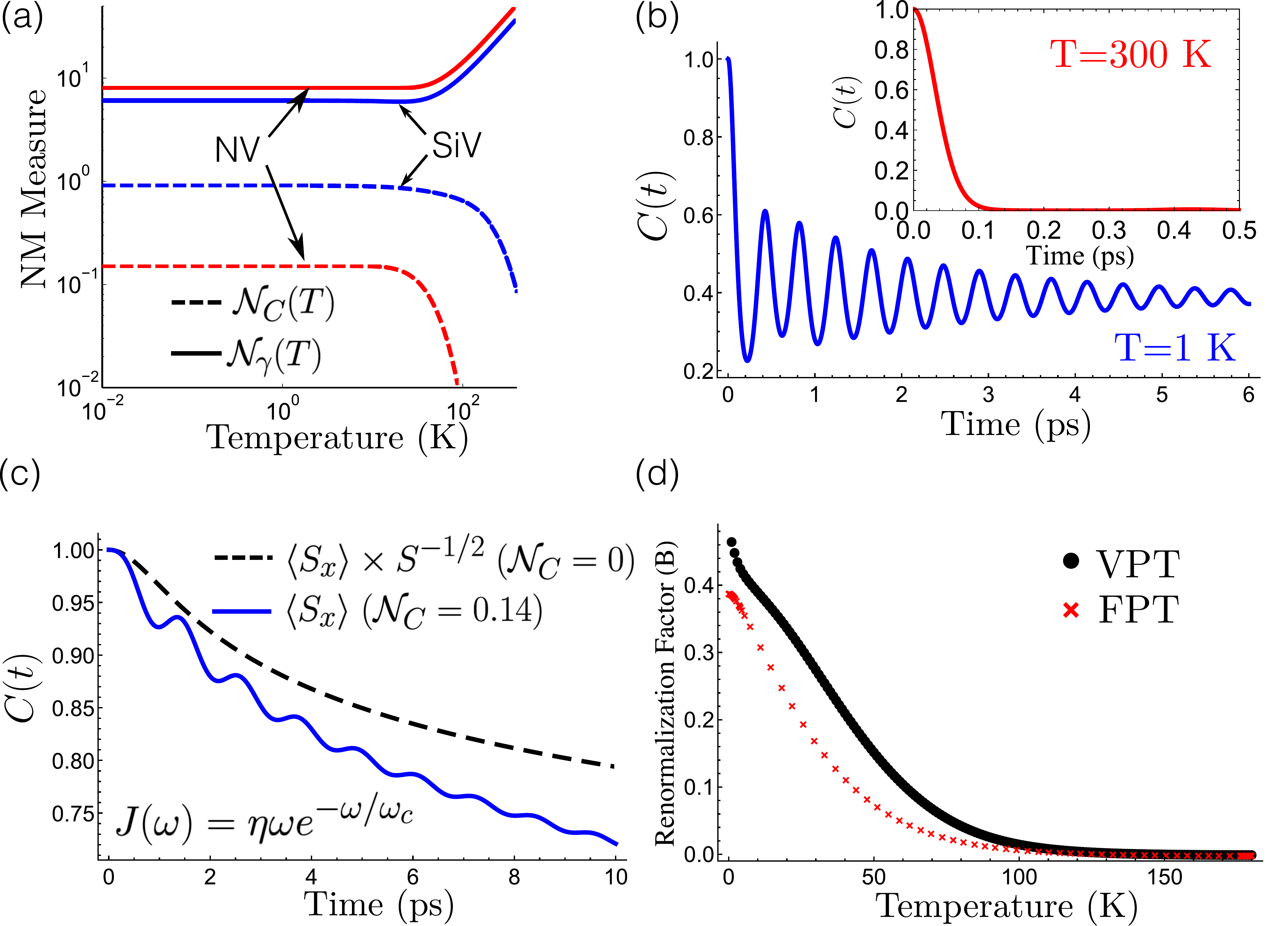}
\caption{(a) Comparison of the non-Markovian measures $\mathcal{N}_C$ and $\mathit{N}_{\gamma}$ for the SiV$^{-}$ and NV$^{-}$ centers in logarithmic scale and for temperatures ranging from $10$ mK to $300$ K. For the SiV$^{-}$ center the SDF $J(\omega) = J_{bulk}(\omega)+J_{loc1}(\omega)+J_{loc2}(\omega)$ was used and for the NV$^{-}$ center the exact SDF given in Ref.~\cite{Alkauskas}. The NM measure $\mathcal{N}_{\gamma}(T)  = \int_{t}^{t'} \left(\vert\gamma (\tau) \vert -\gamma (\tau)\right)\, d\tau$ was calculated using $t=0$ and a sufficient large time $t'= 300$ ps. (b) Coherence function $C(t)$ reveals a back-flow of quantum information at low temperature while at room temperature it monotonically decreases. (c) Filtering function $S^{-1/2}$ allows to extend our measure beyond incoherent operations. We use the values $\eta = 1/600$ and $\omega_{c} = 1$ THz for the sub-Ohmic spectral density function $J(\omega) = \eta \omega \exp(-\omega/\omega_c)$. (d) Renormalization factor $B$ for the SiV$^{-}$ center using the full polaron transformation (FPT) and the variational polaron transformation (VPT) for $\Omega = 0.6$ GHz and $\Delta = \Omega/2$.}
\label{fig:Figure3}
\end{figure}

In Figure~\ref{fig:Figure3}-(a) we show the temperature dependence of $\mathcal{N}_C(T)$ (dashed lines) and $\mathcal{N}_{\gamma}(T)$ (solid lines) for both color centers in diamond (NV$^{-}$ and SiV$^{-}$). The system was prepared in the initial pure state $\rho_0 =|\psi(0)\rangle \langle \psi(0)|$, where $\ket{\psi(0)}=\left(\ket{e}+\ket{g} \right) /\sqrt{2}$, and hereafter we will only focus on this condition. It is interesting that both measures are almost constant at low temperatures, but above $T\approx 100$ K, $\mathcal{N}_{\gamma}(T)$ starts to increase linearly with temperature, while in contrast, $\mathcal{N}_C$ goes to zero. The former can be explained by noticing that $\gamma_{loc1}$ increases with temperature, and would lead to the conclusion that the bath becomes more NM with increasing temperature. For the latter, we plotted the time evolution of the Coherence in Figure~\ref{fig:Figure3}-(b). Here one immediately sees that at high temperature the NM disappears, as one would expect from the coupling to bulk phonons. To shed more light on this matter, the key is to look at the unusual and complex spectral density function of these systems. Onnone hand, at low temperatures the dynamics is ruled out by the SDF $J(\omega)$, since all phonons are frozen out. For this reason, the quasi-localized phonon has the leading contribution to NM, in agreement with the remarks given in Ref.~\cite{Vasile} for an engineered reservoir. On the other hand, at high temperatures, low-frequency (acoustic) phonons are dominant ($\gamma^{\uparrow}(t)\sim \int_{0}^{\infty}J(\omega)/(\hbar \omega)^2 \sin \omega t \,d\omega$), and therefore the reservoir can be modeled by a super-Ohmic spectrum. In this scenario, NM is highly suppressed~\cite{Haikka}.   \par

To further support the physical consistency of our coherence measure,  we compare now $\mathcal{N}_C$ with the well-known measure proposed by Breuer, Laine, and Piilo ($\mathcal{N}_{BLP}$) \cite{Breuer}, where the figure of merit is the trace distance between two quantum states $D(t) = \mbox{tr}|\rho_1-\rho_2|/2$, (see Appendix \ref{Appendix_BLP} for more details). Both measures have a similar behavior as a function of temperature, see Figure \ref{fig:Figure7}, although $\mathcal{N}_C$ neither requires a maximization over the set of initial states ($\rho_1(0)$ and $\rho_2(0)$) nor a full tomography of the density matrix. Moreover, it can be easily computed and experimentally measured through a system observable such as $\langle \sigma_x \rangle$. Also, it does not involve an auxiliary system (in contrast to the case of quantum mutual information measure \cite{Luo}) and reaches its maximum when $\mathcal{N}_C= 1$ (supported by the results in Ref.~\cite{Chruscinski} for the decay rate). \par

\subsection{Non-markovianity measure for coherent maps} \label{NM-Coherent-Maps}
In the current and many other settings of interest, the Rabi-frequency $\Omega$ in 
Hamiltonian~\eqref{Hamiltonian_Spin_Boson_Model} is controlled by an external drive and can be switched off. However, this might not be the case for other systems, where a non-vanishing coupling can induce oscillations between the states $|g\rangle$ and $|e\rangle$. As already pointed out in previous works \cite{Chanda, Passos}, this induces as well oscillations in the Coherence, leading to a false positive when detecting NM. Therefore, we propose here an experimental sequence to specifically filter the dynamics corresponding to the $(\Omega/2)\sigma_x$ component. This procedure is first demonstrated for the weak-coupling regime, where we can derive an exact master equation for $\rho_s(t)$, as detailed in Appendix~\ref{Appendix_WeakCoupling}. The resulting time evolution for the off-diagonal elements is given by 
$\rho_{eg}(t) = \rho_{eg}(0)\mbox{exp}(-F_0(t)+F(t)/2)$, where $F_0(t)$ reduces to $\Gamma(t)$ for $\Omega=0$. The function $F(t)$ contains contributions, which are solely related to coherent population oscillations, and therefore must be cancelled. This can be done by using the fact that the quantity $S=(1/2)(\langle S_z \rangle_{(\rho_{ee}=1)}-\langle S_z \rangle_{(\rho_{gg}=1)})= \mbox{exp}(F(t))$ (as given in Eq.~\eqref{S}) depends on $F(t)$ only. Here the subscripts indicate that the average dynamics of $\langle S_z\rangle$ is calculated starting from the initial excited ($e$) and ground ($g$) eigenstates of the Hamiltonian $H_s = -(\Delta /2)\sigma_z + (\Omega/2)\sigma_x$, respectively. The operators $S_i$ are the usual Pauli matrices written in the system eigenbasis. In summary, we perform the following three steps: $(1)$ perform a measurement of $\langle S_z \rangle$ for the initial condition $\rho_{gg}(0)=1$, $(2)$ repeat the measurement for $\rho_{ee}(0)=1$, $(3)$ calculate the Coherence by performing a measurement of $\langle S_x \rangle$ for the initial pure state $|\varphi(0)\rangle = (|1\rangle + |2\rangle)/\sqrt{2}$, and multiply the outcome by $S^{-1/2}$, where $|1\rangle$ and $|2\rangle$ are defined in Appendix~\ref{Appendix_WeakCoupling}. From this procedure we define a renormalized coherence $\tilde{C}(t) = C(t)\times S^{-1/2}$ which can be used instead of $C(t)$ in Eq.~\eqref{NM_measure}. Note that this sequence is unbiased with the spectral density function and it takes advantage of the dynamics induced by the $\sigma_x$ component. \par

For further illustration, we plotted in Figure~\ref{fig:Figure3}-(c) the coherence calculated with and without the filtering function $S^{-1/2}$ for the SDF $J(\omega)=\eta\omega\exp(-\omega/\omega_{c})$, which is well-know to lead to Markovian dynamics at arbitrary temperaurtures~\cite{Haikka,Fanchini2010}. First, because of the presence of $\sigma_x$-term in Hamiltonian~\eqref{Hamiltonian_Spin_Boson_Model}, $C(t)$ shows an oscillation that the measure $\mathcal{N}_{C}=0.14$ detects as non-Markovianity. Secondly, our filter function eliminates the oscillation, and we obtained $\mathcal{N}_{\tilde{C}}=0$ for $\tilde{C}(t)$, as expected for this SDF. Our sequence takes advantage of the weak-coupling between the two-level system and phonons, and gets the information from $\sigma_x$ and uses it to cancel its contribution on the off-diagonal terms. \par

Beyond the weak coupling regime the generalized polaron transformation offers the possibility to apply a similar procedure for strong system-bath interactions~\cite{Ahsan2011,Lukin2013}. This transformation is defined as $H' = e^{S_1}He^{-S_1}$ with $S_1 = \sigma_z \sum_{k}(f_k /\omega_k)(b_{k}^{\dagger}-b_k)$ and leads to $H' = H_0 + H_{ph} + V$, where $H_0 = -(\Delta /2)\sigma_z + (\Omega_R /2)\sigma_x$ is the system Hamiltonian (neglecting the polaron shift), $H_{ph} = \sum_{k}\omega_k b_k^{\dagger}b_k$ is the phonon Hamiltonian, $V = \sigma_x V_x + \sigma_y V_y + \sigma_z V_z$ is the interaction Hamiltonian (see Eqs.~\eqref{Vx}-\eqref{Vz} in Appendix~\ref{Appendix_Strong_Coupling}) and $\Omega_R = B \Omega$, where $B$ is the renormalization factor~\cite{Ahsan2011,Lukin2013}, see 
Appendix~\ref{Appendix_Strong_Coupling}. In Figure~\ref{fig:Figure3}-(d) we plotted the renormalization factor $B$ for the full polaron transformation (FPT) ($f_k = g_k$) and the variational polaron transformation (VPT) ($f_k = F(\omega_k)g_k$), where $F(\omega_k) = [1+ \Omega_R^2/(\omega_k \omega_0)\mbox{coth}\left(\beta \omega_k /2 \right) \mbox{tanh}\left(\beta \omega_0 / 2 \right)]^{-1}$. We observed that for not too large values of $\Omega$, the renormalized term $(\Omega_R/2)\sigma_x$ has a negligible effect in the open dynamics above a temperature $T \approx 120$ K. This temperature is very close to the temperature associated with the quasi-localized phonon mode $T_{loc} = \hbar \omega_{loc}/k_B \approx 116$ K. Therefore, oscillations induced by $\sigma_x$ on the coherence will not be observed at high temperatures \cite{Lukin2013}. \par

\section{Conclusions}

In summary, we studied a simple measure to quantify the degree of Non-Markovianity (NM) that takes advantage of the pure dephasing dynamics and measures the back-flow of information from the environment to the system. We compare it with other established measures of NM~\cite{Breuer,Hall, Rivas}, and observed that it is well behaved both at low and high temperatures. Also, it is easy to compute and experimentally accessible through Ramsey spectroscopy. Moreover, we studied the dynamics of SiV$^-$ and NV$^-$ centers due to the vibrations of the diamond lattice and found that the competition of acoustic and quasi-localized phonon modes give rise to a NM dynamics with a rich thermal dependence. As a consequence, different measures of NM exhibit opposite dependence at high temperature, suggesting that the power of each measure will rely on how it is linked to a specific application. Furthermore,  we extended this NM measure from incoherent (independent spin-boson model) to coherent dynamical maps (general spin-boson model), where for the weak coupling we found that following a specific experimental sequence one can filter out the undesired contribution of the $\sigma_x$-term, while for the strong coupling regime there is a temperature $T \approx 120$ K above which this term is negligible. 

\section{acknowledgments} We are grateful to Felipe Fanchini for fruitful discussions about non-Markovianity. AN acknowledges financial support from Universidad Mayor through the Postdoctoral fellowship. JRM acknowledges support from Fondecyt Regular No. 1180673 and AFOSR FA9550-18-1-0513. PR acknowledges support from the Austrian Science Fund (FWF) through Grant No. P 31701-N27. RC acknowledge financial support from Fondecyt Iniciaci\'on No. 11180143. 

\appendix

\section{Exact reduced dynamics for the independent spin-boson model}
In this section we introduce the exact master equation associated with the reduced two-level system (TLS) of the independent spin-boson Hamiltonian given by
\begin{equation}
H  = {1 \over 2}\omega_0 \sigma_z + \sum_{k} \omega_k a_{k}^{\dagger}a_k + \sigma_z \sum_{k} g_k \left( a_k +a^{\dagger}_k \right), 
\end{equation}
where $\omega_0$ is the bare frequency of the TLS, $\omega_k$ are the phonons frequencies and $g_{k}$ are the electron-phonon coupling constants. First, we consider the Born approximation for the initial state $\rho(0) = \rho_s(0) \otimes \rho_{ph}$, where $\rho_{ph} = e^{-\beta H_{ph}}/\mbox{Tr}\{e^{-\beta H_{ph}}\}$ is a thermal bosonic state, with $\beta = (k_B T)^{-1}$, 
$H_{ph} = \sum_{k} \omega_k a_{k}^{\dagger}a_k$ and $\rho_s(t) = \mbox{Tr}_{ph}(\rho(t))$. In the Heisenberg picture, the exact equations of motion for the expectation values $\langle \sigma_{\pm}(t) \rangle$ are given by \cite{Luczka}
\begin{equation}
{d\langle \sigma_{\pm}(t) \rangle  \over dt} = \left[\pm {i\omega_0 \over 2} - \gamma(t) \right]\langle \sigma_{\pm}(t) \rangle,
\label{A1}
\end{equation}
where $\sigma_{\pm} = (\sigma_x \pm i\sigma_y)/2$, $\langle \sigma_{\pm}(t) \rangle  = \mbox{Tr}_s\left( \sigma_{\pm}(t) \rho_s(0) \right)$, $\gamma(t)$ is the dephasing rate introduced in Eq.~\eqref{DephasingRate}, $T$ is the reservoir temperature, $k_B$ is the Boltzmann constant, and $ J(\omega) = \sum_{k} |g_k|^2\delta(\omega - \omega_k)$ is the spectral density function of the reservoir. In general, the density matrix of a two-level system can be written as $\rho_s(t) = (1 / 2) \left(\mathds{1} + \langle \vec{\sigma}(t) \rangle \cdot \vec{\sigma} \right)$, where $\vec{\sigma} = (\sigma_x, \sigma_y, \sigma_z)$, $\mathds{1}$ is the identity matrix, and $\sigma_i$ are the usual Pauli matrices. Using the last decomposition and Eq.~\eqref{A1}, we derived the following equation
\begin{equation}
{d\rho_s(t) \over dt}  = -i[H_s,\rho_s(t)] - {\gamma(t) \over 2}\left(\langle \sigma_x(t) \rangle \sigma_x+ \langle \sigma_y(t) \rangle \sigma_y \right),
\end{equation}
where we have used $\sigma_z(t) = \sigma_z(0)$ (constant of motion) due to the fact that $[H_s,\sigma_z] = 0$ for 
$H_s = (\omega_0 /2)\sigma_z$. Using the property $\sigma_i \sigma_ j = \delta_{ij} \mathds{1} + i\epsilon_{ijk} \sigma_k$, we obtained the relation
\begin{flalign}
&\sigma_z \rho_s(t) \sigma_z = {1 \over 2}\left(\mathds{1}-\langle \sigma_x(t) \rangle\sigma_x-\langle \sigma_y(t) \rangle\sigma_y+\langle \sigma_z(t) \rangle\sigma_z \right). 
\end{flalign}
From the above relation and $\rho_s(t) = (1 / 2) \left(\mathds{1} + \langle \vec{\sigma}(t) \rangle \cdot \vec{\sigma} \right)$ we get $\langle \sigma_x(t) \rangle \sigma_x+ \langle \sigma_y(t) \rangle \sigma_y = \rho_s(t)-\sigma_z \rho_s(t) \sigma_z$. Therefore, the master equation for $\rho_s(t)$ is given by
\begin{equation}
{d\rho_s(t) \over dt}  = -i[H_s,\rho_s(t)] - {\gamma(t) \over 2}\left(\rho_s(t)-\sigma_z \rho_s(t) \sigma_z\right).
\end{equation}
By neglecting the first term we recover the dynamics in the interaction picture (see Eq.~(2) of the main text).

\section{Dephasing rate induced by strong interactions with quasi-localized phonons}\label{Appendix_gamma}

The dephasing rate associated with the strong interaction with a quasi-localized phonon mode is given by
\begin{equation}
\gamma_{loc1}(t)  = {J_0 \Gamma \over 2}\int_{0}^{\infty} {\omega^2  \coth\left( {\omega \over 2 k_B T}\right) \sin(\omega t)  \over \left({\omega \over \omega_{loc}} +1\right)^{2}(\omega-\omega_{loc})^2 + (\Gamma/2)^2}\, d\omega.
\end{equation}
This integral can be solved analytically, however, we show next a method to obtain a good approximation that gives us a better understanding of the effect of the width $\Gamma$, the frequency of the quasi-localized phonon $\omega_{loc}$, the amplitude $J_0$, and temperature $T$. Using the change of variable $u = \omega-\omega_{loc}$ and extending the lower limit of the integration to $-\infty$ (assuming $\omega_{loc} \gg 1$), we obtain 
\begin{eqnarray}
\gamma_{loc1}(t)  & \approx & J_0   \cos(\omega_{loc}t)\int_{-\infty}^{\infty} f(u)\sin (ut) \, du \nonumber\\ &+& J_0 \sin(\omega_{loc}t)\int_{-\infty}^{\infty} f(u)\cos (ut) \, du \label{Integral} 
\end{eqnarray}
where
\begin{equation}
f(u) = {\left(u+\omega_{loc}\right)^2 \coth\left( {u+\omega_{loc}  \over  2 k_B T}\right) \over \left(u / \omega_{loc}+2 \right)^2 } {\Gamma/2 \over u^2 + \left(\Gamma/2 \right)^2}.
\end{equation}
The main contribution in both integrals given in Eq.~\eqref{Integral} comes from the narrow Lorentzian function $L(u) = (\Gamma/2)/(u^2+(\Gamma/2)^2)$ around the value $u=0$ (main peak around $\omega = \omega_{loc}$ for $J(\omega)$, see Figure~\ref{fig:Figure1}). Using the approximation
\begin{equation}
f(u) \approx {1 \over 4}\omega_{loc}^2 \coth\left({\omega_{loc}\over 2k_B T}\right){\Gamma/2 \over u^2 + \left(\Gamma/2 \right)^2},
\end{equation}
and the symmetry consideration $\int_{-\infty}^{\infty}f(u)\sin(ut)\, du = 0$, we obtain
\begin{equation}
\gamma_{loc1}(t)  \approx  {\pi  J_0 \omega_{loc}^2\over 4} \coth\left( {\omega_{loc} \over 2 k_B T}\right) \sin(\omega_{loc}t)e^{-\Gamma t/2},
\end{equation}
which corresponds to a damped periodic oscillation, where $\omega_{loc}$ is the frequency, $\Gamma/2$ is the decay rate and both temperature $T$ and strength $J_0$ determine the maximum amplitude of the oscillations. At zero temperature, $\coth(\hbar \omega /2k_B T) = 1$, and therefore, we recover the expression given in Eq.~\eqref{gamma_loc1_lowT}. At high temperatures, $k_B T \gg \hbar \omega_{loc}$, we have $\coth(\hbar\omega/2k_BT) \approx 2k_BT/\hbar \omega$, and then
\begin{eqnarray}
\gamma_{loc1}^{\uparrow}(t) \approx \left({2 k_B T \over \omega_{loc}} \right) \gamma_{loc1}^{\downarrow}(t).
\end{eqnarray}

\section{Dephasing Rate for an NV$^-$ center}\label{Appendix_NV}

The dephasing rate $\gamma_{NV}(t)$ is calculated from Eq.~\eqref{DephasingRate} with the spectral density function given numerically in Ref.~\cite{Alkauskas}, and illustrated in Figure~\ref{fig:Figure1}.

\begin{figure}[ht!]
	\centering
	\includegraphics[width = .8 \linewidth]{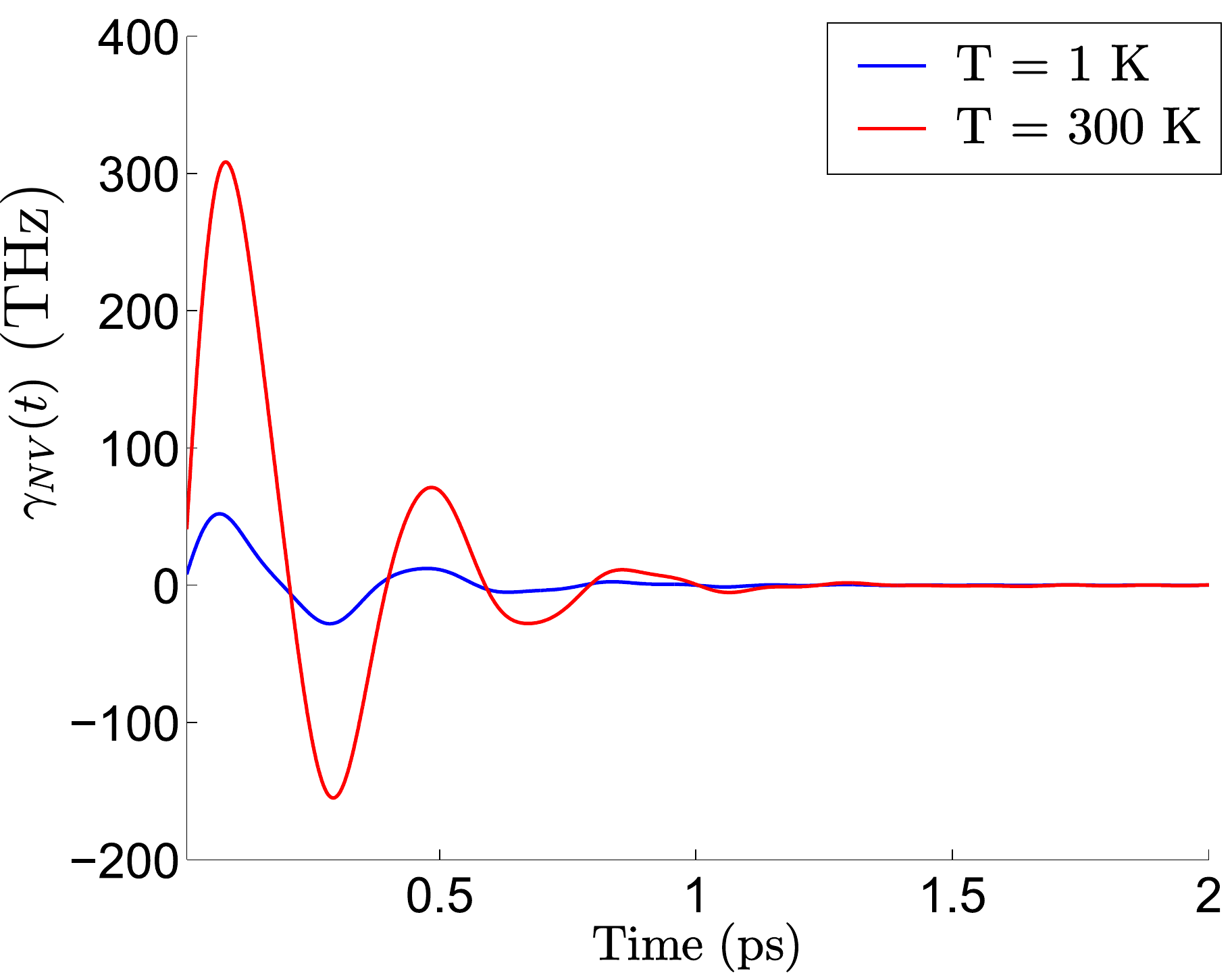}
	\caption{Dephasing rate for the NV$^{-}$ center in diamond for temperatures $T = 1$ K and $T = 300$ K. The period of the oscillations are given approximately by $2\pi / \omega \approx 0.4$ ps, where $\omega = 15.7$ THz is the frequency of the main peak of the spectral density function of the NV$^{-}$ center.}
	\label{fig:FigureNV}
\end{figure}

\section{Measures of the degree of non-Markovianity}\label{Appendix_BLP}
\begin{figure}[ht!]
	\centering
	\includegraphics[width = 0.8 \linewidth]{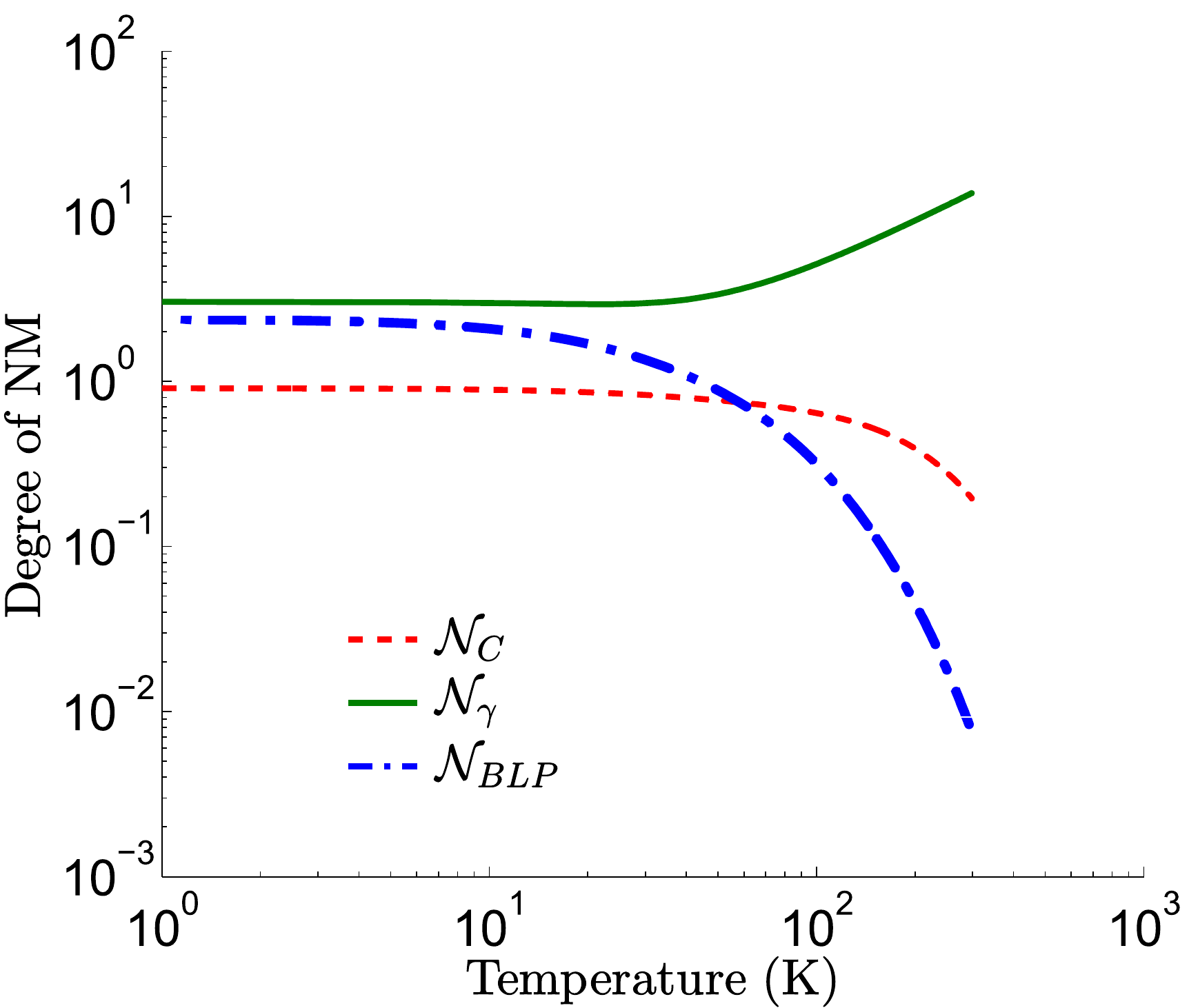}
	\caption{Comparison between different measures of non-Markovianity as a function of the reservoir temperature.}
	\label{fig:Figure7}
\end{figure}

For comparison, we introduce now the Breuer-Laine-Piilo (BLP) measure \cite{Breuer}, which is based on the trace distance 
\begin{equation}
\mathcal{N}_{BLP} = \int_{dD/dt>0}^{} {d D \over d \tau} \, d\tau, \quad \quad D(t) = {1 \over 2}\mbox{Tr}|\rho_1(t)-\rho_2(t)|,
\end{equation}  
where $|A| = \sqrt{A^{\dagger}A}$. The BLP measure is plotted in Figure.~\ref{fig:Figure7} (dotted line) for the initial states  
\begin{equation}
\rho_1(0) = {1 \over 2}\left(\begin{array}{ll}
1 & 1 \\
1 & 1 
\end{array} \right),  \quad \quad 
\rho_2(0) = {1 \over 2}\left(\begin{array}{rr}
1 & -1 \\
-1 & 1 
\end{array} \right).                            
\end{equation} 
In addition, we noted that the Rivas-Huelga-Plenio (RHP) measure is given by the relation $\mathcal{N}_{\gamma} = (d/2)\mathcal{N}_{RHP}$~\cite{Rivas}, where $d=2$ is the dimension of the Hilbert space in our case (two-level system), and therefore $\mathcal{N}_{\gamma}$ and $\mathcal{N}_{RHP}$ are identical. \par

Notice that $\mathcal{N}_{\gamma}$ (solid line) shows a thermal dependence that is based on the dephasing rate $\gamma(t)$ given in Eq.~\eqref{DephasingRate}, and it has a strictly increasing behavior leading to a large degree of NM when the temperature is above some critical value $T\sim 30$ K. However, we observed that $\mathcal{N}_{BLP}$ shows a strictly decreasing behavior leading to a small degree of NM at room temperature. Finally, $\mathcal{N}_C$ (dashed-line) shows a thermal dependence that is similar to $\mathcal{N}_{BLP}$, as one would expect since both NM measures are based on the back-flow of quantum information.

\section{Weak coupling model}\label{Appendix_WeakCoupling}
For a better understanding of the dynamics of the system when it is weakly coupled to the environment, we traced over the reservoir in Eq.~\eqref{Hamiltonian_Spin_Boson_Model}, assuming that the reservoir is in a thermal state. After the Born and the secular approximations, one gets the following master equation~\cite{Laine},
\begin{equation}
\dot{\rho_s}(t)=-i[H_s,\rho_s(t)] + D\rho_s(t),
\end{equation}
where $H_s = (\omega_0/2) S_z$,  $\omega_0=\sqrt{\Delta^2 + \Omega^2}$, and $S_z=|1\rangle\langle 1|-|2\rangle\langle 2|$, being 
\begin{eqnarray}
|1\rangle &=& \dfrac{\Omega|e\rangle + (\omega_0+\Delta)|g\rangle}{\sqrt{2\omega_0(\omega_0+\Delta)}}, \\
|2\rangle &=& \dfrac{-\Omega |e\rangle + (\omega_0-\Delta)|g\rangle}{\sqrt{2\omega_0(\omega_0-\Delta)}},
\end{eqnarray}
the system eigenstates. Notice that the Lamb shift Hamiltonian has been omitted since it is negligible according our numerical simulations. Furthermore, the Lindblandian is given by,
\begin{equation}
D\rho_s(t)=\sum_{\xi= 0,\pm\omega_0 } \frac{\gamma_{\xi}(t)}{2}\left(L_\xi\rho_s(t)L^\dagger_\xi-\frac{1}{2}\lbrace L^\dagger_\xi L_\xi,\rho_s(t) \rbrace \right).
\end{equation}
Here, $L_0 = (\Delta/\omega_0) S_z$, $L_{+\omega_0}=(\Omega/\omega_0) S_-$ and $L_{-\omega_0}=(\Omega/\omega_0) S_+$, where $S_+ = |1\rangle\langle 2|$ and $S_-=|2\rangle\langle 1|$ are the raising and lowering operators in the system eigenbasis. Moreover, for arbitrary temperatures, we have,
\begin{eqnarray}
\gamma_\xi (t) &=& \int_0^\infty J(\omega) \left[n(\omega){\sin(\omega+\xi)t \over \omega+\xi} \right. \nonumber \\
               & & \left. + [n(\omega)+1]{\sin(\omega-\xi)t \over \omega-\xi}\right] d\omega,	
\end{eqnarray}
where $n(\omega) = [\mbox{exp}(\hbar \omega/k_B T)-1]^{-1}$ is the mean number of phonons at thermal equilibrium and $J(\omega)$ is the spectral density function introduced in Eq.~\eqref{DephasingRate}. One can immediately notice that for $\Omega=0$, we recover the dynamics in Eq.~\eqref{ME}. Since the Hamiltonian evolution $\sim [S_z,\rho_s(t)]$ does not affect the coherence, we will only focus on the dynamics induced by $D\rho_s(t)$. The time evolution of the density operator is given by,
\begin{eqnarray}
\dot{\rho}_{ee}(t)&=&\frac{\Omega^2}{\omega_0^2}\left(\gamma_{-\omega_0}(t)-(\gamma_{-\omega_0}(t)+\gamma_{\omega_0}(t))\right)\rho_{ee}(t), \\
\dot{\rho}_{eg}(t)&=&-2\frac{\Delta^2}{\omega_0^2}\gamma_0(t) \rho_{eg}(t) \nonumber \\ &-& \left(\frac{\Omega^2}{2\omega_0^2}(\gamma_{-\omega_0}(t)+\gamma_{\omega_0}(t))\right)\rho_{eg}(t),
\end{eqnarray}
where $\rho_{ee}(t) = \langle 1| \rho(t) |1\rangle$, $\rho_{gg}(t) = \langle 2| \rho(t) |2\rangle$ and 
$\rho_{eg}(t) = \langle 1| \rho(t) |2\rangle$. The above equations can be formally integrated, 
\begin{eqnarray}
\rho_{ee}(t)&=&\rho_{ee}(0)e^{F(t)} + e^{F(t)}\int_0^t  g(t_1)e^{-F(t_1)} \, dt_1, \\
\rho_{eg}(t)&=& \rho_{eg}(0)e^{-F_0(t)}e^{F(t)/2}, \label{rho_eg}
\end{eqnarray}
where $F(t)=-(\Omega/\omega_0)^2\int_0^t (\gamma_{-\omega_0}(t_1)+ \gamma_{\omega_0}(t_1)) \, dt_1$, $g(t) = (\Omega/\omega_0)^2\gamma_{-\omega_0}(t)$, and $F_0(t) = 2(\Delta/\omega_0)^2\int_0^t \gamma_0(t_1) \,dt_1$. Notice that $F_0(t)$ corresponds to the dynamics induced by $S_z$, that does not generate coherence. Only the contribution of $F(t)$ in Eq.~\eqref{rho_eg} will lead to a false non-Markovianity, and therefore we will cancel it. To do so, we first calculate the expectation values of $S_z$ for different initial conditions.
\begin{eqnarray}
 	\langle S_z \rangle_{(\rho_{ee}=1)}&=& 2 e^{F(t)} + \langle S_z \rangle_{(\rho_{ee}=1)},    \\
 	\langle S_z \rangle_{(\rho_{gg}=1)}&=& 2 e^{F(t)}\int_{0}^{t} g(t_1)e^{-F(t_1)}\,dt_1 -1.
\end{eqnarray}  
From these results it is straight forward to obtain that
\begin{equation} \label{S}
S = \frac{\langle S_z \rangle_{(\rho_{ee}=1)}-\langle S_z \rangle_{(\rho_{gg}=1)}}{2}=e^{F(t)}.
\end{equation}
Furthermore, we have that $\langle S_x \rangle = 2\rho_{eg}$, given that $\rho_{eg}\in\mathbb{R}$ and $S_x = S_+ + S_-$. Hence, for our generalize measure of NM, the Coherence will be defined as $\tilde{C}(t)=\langle S_x \rangle\times S^{-1/2}=2\rho_{eg}(0)e^{-F_0(t)}$, where for an incoherent dynamics $S=1$. Finally, one can observe that this expression for the Coherence has no contribution from $(\Omega/2)\sigma_x$ in Eq.~\eqref{Hamiltonian_Spin_Boson_Model} (as opposed to the case where $C(t)$ is calculated directly from Eq.~\eqref{rho_eg} and even more, we have not done any assumption about the spectral density function $J(\omega)$, which makes this approach quite general. \par

\section{Strong coupling model}\label{Appendix_Strong_Coupling}
In order to study strong interactions between a two-level system and its phononic environment we introduce the following general polaron transformation~\cite{Lukin2013,Ahsan2011} ($\hbar = 1$)
\begin{equation}
H' = e^{S_1} H e^{-S_1}, \quad \quad S_1 = \sigma_z \sum_{k} {f_k \over \omega_k}(b_k^{\dagger}-b_k),
\end{equation}
where $f_k = g_k$ corresponds to the the full polaron transformation (FPT). If we apply the above transformation on the general spin-boson Hamiltonian $H = -(\Delta / 2) \sigma_z + (\Omega / 2) \sigma_x +\sum_{k}\omega_k b_k^{\dagger} b_k + \sigma_z \sum_{k} g_k(b_k^{\dagger}+b_k)$, we obtain
\begin{equation}
H' = -{\Delta \over 2} \sigma_z + {\Omega_R \over 2} \sigma_x + \sum_{k}{f_k \over \omega_k}\left(f_k-2g_k \right)+\sum_{k}\omega_k b_k^{\dagger} b_k + V.
\end{equation}
The interaction Hamiltonian is given by $V = \sigma_x V_x + \sigma_y V_y + \sigma_z V_z$, where 
\begin{eqnarray}
V_x &=& {\Omega \over 4}\left(B_+ + B_-  - 2B\right), \label{Vx}\\
V_y &=& {\Omega \over 4i}\left(B_- - B_+\right), \label{Vy} \\
V_z &=& \sum_{k}(g_k-f_k)(b_{k}^{\dagger}+b_k). \label{Vz} 
\end{eqnarray}
The bath operators are defined as 
\begin{eqnarray}
B_{\pm} &=& \mbox{exp}\left[\pm 2 \sum_{k}{f_k \over \omega_k}(b_{k}^{\dagger}-b_k) \right],\\
B &=& \langle B_{\pm} \rangle_{ph} = \mbox{exp}\left[- 2 \sum_{k}{f_k^2 \over \omega_k^2}\mbox{coth}\left({\beta \omega_k \over 2 }\right) \right],
\end{eqnarray}
with $\beta = (k_B T)^{-1}$ and the expectation value is calculated using the thermal phonon state $\rho_{ph}=\mbox{exp}(-\beta H_{ph})/\mbox{Tr}\{\mbox{exp}(-\beta H_{ph})\}$. To determine the optimal values of $f_k$ for the variational polaron transformation (VPT) it is necessary to minimize the free energy 
\begin{equation}
A_B = -{1 \over \beta}\mbox{ln} \mbox{Tr}_{A+B}\{e^{-\beta H_0}\} + \langle V \rangle_{H_0},
\end{equation}
where $H_0 = H_{A} + H_{B}$ is the non-interacting Hamiltonian with $H_A  = -(\Delta / 2) \sigma_z + (\Omega_R / 2) \sigma_x+ \sum_{k}(f_k / \omega_k)\left(f_k-2g_k \right)$ and $H_B = \sum_{k}\omega_k b_k^{\dagger} b_k$. Using $\langle V \rangle_{H_0}=0$ and the condition $d A_B/ d f_k =0$ we obtain that $f_k = g_k F(\omega_k)$, where
\begin{eqnarray}
F(\omega_k) &=& \left[1+ {\Omega_R^2 \over \omega_k \omega_0}\mbox{coth}\left({\beta \omega_k \over 2} \right)
\mbox{tanh}\left({\beta \omega_0 \over 2} \right) \right]^{-1}, \label{F-function} \\
\Omega_R &=& \Omega B, \label{RenormalizationConstant}
\end{eqnarray}
and $\omega_0 = \sqrt{\Delta^2+\Omega_R^2}$. In the continuum limit, the renormalization factor $B$ is given by
\begin{equation}
B = \mbox{exp}\left[- 2 \int_{0}^{\infty}{J(\omega) \over \omega^2}F^2(\omega)\mbox{coth}\left({\beta \omega \over 2}\right) \, d\omega\right]. \label{RenormalizationFactor}
\end{equation}
This renormalization factor depends on the shape of the SDF $J(\omega)=\sum_{k}|g_k|^2\delta(\omega-\omega_k)$, the reservoir temperature $T$, and must be calculated from self-consistency between Eqs.~\eqref{F-function}-\eqref{RenormalizationFactor}.

\section{Role of different spectral density function on the Coherence}

\begin{figure}[ht!]
\centering
\includegraphics[width = 0.8 \linewidth]{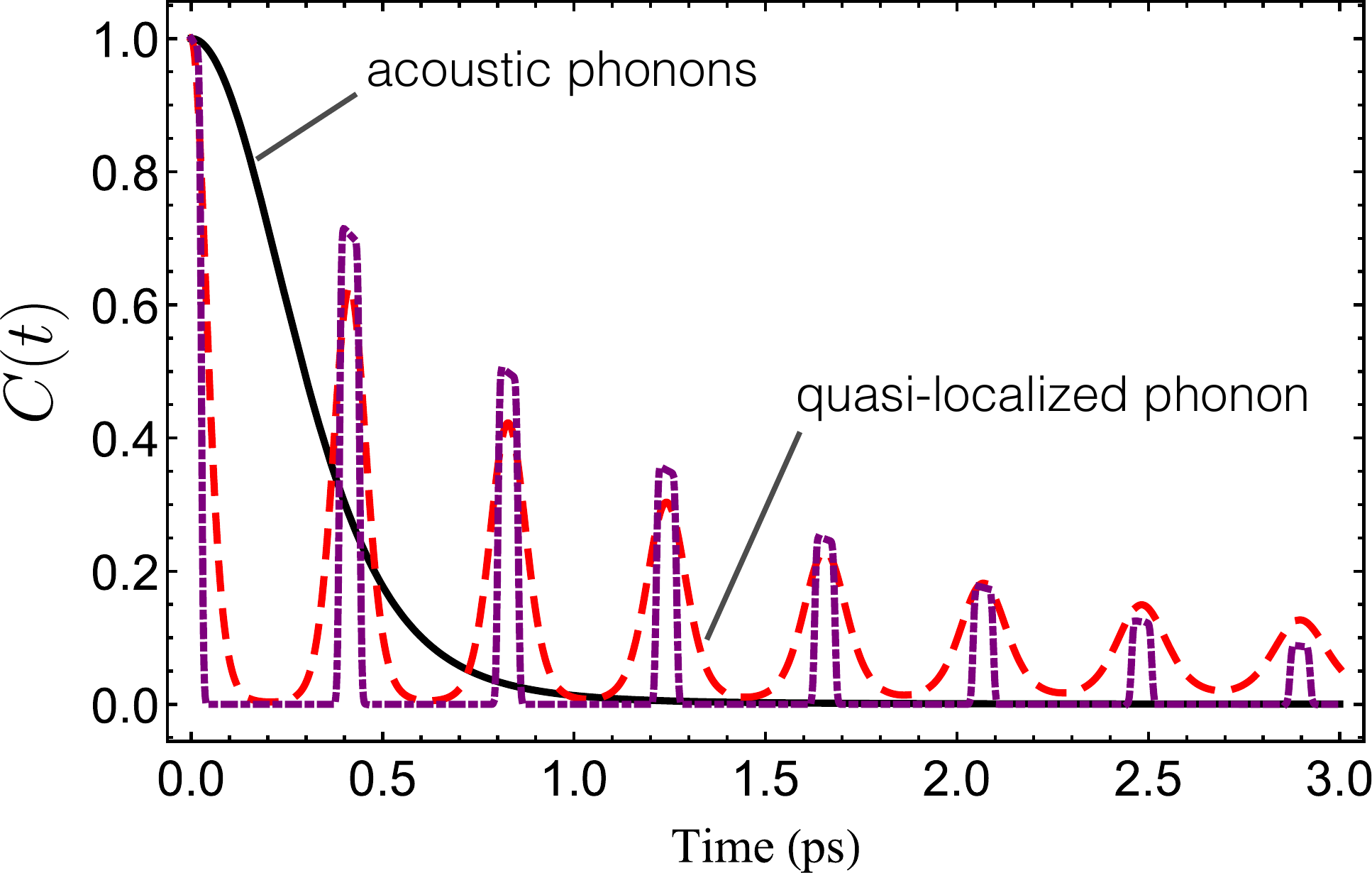}
\caption{Behavior of the coherence function $C(t)$ at $T=300$ K. Solid line (evaluated for $J_{bulk}(\omega)$ only) evidences that bulk phonons have the main contribution to the dynamics. Moreover, dashed line (evaluated for $J_{loc1}(\omega)$ only), evidences that quasi-localized phonon modes can be roughly reproduced with an interaction with a coherent state and a phenomenological decay (dot-dashed).}
\label{fig:Figure6}
\end{figure}

From Figure~\ref{fig:Figure3}-(b), it is easy to see that Coherence $C(t)$ obeys two different regimes separated in temperature. At low temperature, the time evolution of $C(t)$ has contributions from of all the spectral density functions in Eqs.~\eqref{Jbulk}-\eqref{Jloc2}, however the strong oscillation in the main plot evidences that $\gamma_{loc1}$ is the leading contribution. Even when one would expect $\gamma_{loc1}$ to hold as the leading term in the dynamics at high temperatures, see Figure~\ref{fig:Figure2}, this is not the case. In Figure~\ref{fig:Figure6} we showed $C(t)$ for two particular decay rates, namely $\gamma_{bulk}(t)$ (solid) and $\gamma_{loc1}(t)$ (dashed), at $300$ K. It is remarkable that $\gamma_{bulk}(t)$ reproduces the behavior (in terms of Non-Markovianity) depicted in the inset of Figure~\ref{fig:Figure3}-(b), despite that it decays slower due to the absence of the other decay rates. This outcome supports the statement that only bulk phonons are relevant at high temperature~\cite{Goldman2015,Goldman2017}. In contrast, $\gamma_{loc1}(t)$ shows a very interesting dynamics as well. \par

In particular, similar behavior have been observed in spin-echo spectroscopy for a single NV$^-$~\cite{Childress} or an ensemble~\cite{Stanwix} interacting with a natural environment of $^{13}\mbox{C}$ nuclear spins, where the collapses and revivals originate from a coherent interaction with individual proximal nuclear spins, and the revivals decay comes from the interaction with the $^{13}\mbox{C}$ bath. In the same way, this behavior has been observed for a single NV interacting with a mechanical oscillator~\cite{Bennet}. In agreement with these results, we suggest that the interaction with this quasi-localized phonon mode can be though as an interaction with a single mode in a coherent state, with an added phenomenological decay that goes as $\exp(-\Gamma t)$, being $\Gamma = 0.8414$ THz the width of the Lorentzian spectral density function. The composite state after this interaction is given by~\cite{Montenegro}, 
\begin{equation}
\vert\Psi\rangle =(\vert e\rangle\vert\beta_\uparrow\rangle + e^{-2i\lambda\beta\sin (t\omega_{loc})}\vert g\rangle\vert\beta_\downarrow\rangle)/\sqrt{2},
\end{equation}
where $\xi=1-e^{-it\omega_{loc}}$, $\vert\beta_\uparrow\rangle =\vert\beta e^{-it\omega_{loc}} + \lambda\xi\rangle$ ($\vert\beta_\downarrow\rangle = \vert\beta e^{-it\omega_{loc}} - \lambda\xi\rangle$) is the displaced coherent state, $\lambda=(\int_{0}^{\infty} J_{loc1}(\omega))^{1/2} \, d\omega$. Even more, we approximated $\vert \beta\vert^2$ to the thermal occupancy phonon number $n(\omega_{loc}) = [\exp(\hbar\omega_{loc}/k_bT)-1]^{-1}$, given that at $T=300$ K the occupancy is small ($n(\omega_{loc}) \approx 2$). Finally, after tracing out over the coherent state degrees of freedom we are able to calculate $C(t)$. This highly simplified model is capable to capture the collapses, revivals, and the overall decay, as illustrate by the dot-dashed curve.

\bibliographystyle{unsrt}

\end{document}